\documentclass[twocolumn,superscriptaddress,showpacs,floatfix,amsmath,amssymb]{revtex4-1}
\usepackage{xcolor}
\usepackage{graphicx}
\usepackage[dvipdfmx,colorlinks=true,linkcolor=blue,citecolor=blue,urlcolor=blue]{hyperref}
\usepackage[utf8]{inputenc}
\usepackage{float}

\newcommand{\fLi}{${}^6$Li}
\newcommand{\bYb}{${}^{174}$Yb}

\newcommand{\fEr}{${}^{167}$Er}
\newcommand{\cm}{\mathrm{cm}}
\newcommand{\um}{\mu\mathrm{m}}
\newcommand{\nm}{\mathrm{nm}}
\newcommand{\ms}{\mathrm{ms}}
\newcommand{\s}{\mathrm{s}}
\newcommand{\G}{\mathrm{G}}

\newcommand{\Hz}{\mathrm{Hz}}

\newcommand{\uK}{\mu\mathrm{K}}

\newcommand{\lossunit}{10^{-22}~\cm^3\,\s^{-1}}

\begin{document}

\title{Observation of Feshbach resonances in an \texorpdfstring{\fEr-\fLi}{168Er-6Li}\ Fermi-Fermi mixture}

\author{F.~Sch\"{a}fer}
\email{schaefer@scphys.kyoto-u.ac.jp}
\affiliation{Department of Physics, Graduate School of Science, Kyoto University, Kyoto 606-8502, Japan}

\author{Y.~Haruna}
\affiliation{Department of Physics, Graduate School of Science, Kyoto University, Kyoto 606-8502, Japan}

\author{Y.~Takahashi}
\email{takahashi.yoshiro.7v@kyoto-u.ac.jp}
\affiliation{Department of Physics, Graduate School of Science, Kyoto University, Kyoto 606-8502, Japan}

\date{\today}

\begin{abstract}
	We present our experimental investigation of the interspecies Feshbach
	spectrum in a mixture of \fEr($F = 19/2, m_F = -19/2$)-\fLi($F = 1/2, m_F =
	1/2$) atoms in the microkelvin temperature regime. These temperatures are
	achieved by means of sympathetic cooling with \bYb\ as a third species.
	Interspecies Feshbach resonances are then identified by investigation of the
	Er-Li inelastic collisional properties for magnetic fields up to $800\,\G$.
	Numerous narrow resonances as well as six resonances with widths above
	$1\,\G$ could be identified. It is these broader resonances that hold much
	promise for interesting future investigations of, for exmample, novel
	superfluid states and  Efimov states in large mass-imbalanced, all-fermionic
	two-component systems.
\end{abstract}

\maketitle

\section{Introduction}
\label{sec:intro}

Ultracold mixtures of quantum gases are a fascinating tool to study in detail
the intricate physics of many-body problems~\cite{bloch_many-body_2008} with
both high controllability and high precision. The high degree of control
extends to both dimensional control by means of optical
lattices~\cite{bloch_ultracold_2005} and interaction control by use of
Feshbach resonances~\cite{inouye_observation_1998, chin_feshbach_2010}.
Generally speaking, the realm of many-body physics extends far beyond the
field of ultracold atomic physics and serves as an interface to interconnect
the different research disciplines in physics. Amongst many possible examples
the Efimov effect stands out in his broad
applicability~\cite{naidon_efimov_2017}. Originally investigated by Vitaly
Efimov in his work on nuclear three-body problems, his famous discovery of
three-body bound states is now a prime example of universal many-body physics
that can be applied to nearly any field of quantum physics. Physical evidence
for the predicted infinite series of three-body bound states, however,
remained elusive for more than 30 years until first evidences were found in
ultracold quantum gases of Cs~\cite{kraemer_evidence_2006,
zaccanti_observation_2009}. However, due to the large scaling factors involved
in the observation of at least three Efimov states it was not until the advent
of mass-imbalanced mixture experiments that multiple Efimov states could be
observed~\cite{pires_observation_2014, tung_geometric_2014}. These seminal
experiments utilized the large mass-imbalance in the Fermi-boson
\fLi-${}^{133}$Cs mixture to reduce the scaling factor from $22.7$ to just
$4.9$.

Based on these hallmark experiments, our present work strives to expand the
experimental possibilities towards more ``exotic'' Efimov states involving two
heavy Fermions that resonantly couple to a third, light particle. Going beyond
the usual Efimov scenario, these novel trimer-states are predicted to only
occur for mass ratios beyond a threshold value of $13.6$ where the scaling
ratio initially diverges~\cite{naidon_efimov_2017}. However, with increasing
mass imbalance this scaling factor quickly reduces to experimentally feasible
values of less than $10$. We here present first steps towards the experimental
realization of these states with a Fermi-Fermi \fEr-\fLi\ mixture that with a
mass ratio of $27.8$ lies well above the critical value of $13.6$ for the
occurrence of these hitherto unobserved Efimov states. In our experiments we
demonstrate successful cooling of the mixture to microkelvin temperatures and
we probe the system for broad interspecies Feshbach resonances. Such broad
Feshbach resonances are required for the necessary fine-tuned control of the
interspecies interactions over several orders of magnitude.

In addition, the realized mass-imbalanced Fermi-Fermi mixture of \fEr-\fLi\
is, together with tunable interactions, promising for realizing a novel
superfluid state with a spatially varying order parameter, known as a
Fulde-Ferrell-Larkin-Ovchinnikov (FFLO)
state~\cite{fulde_superconductivity_1964, larkin_nonuniform_1965,
radzihovsky_imbalanced_2010, wang_enhancement_2017}, and also other new
behaviors like a Lifshitz point~\cite{gubbels_imbalanced_2013,
gubbels_lifshitz_2009}. Quite recently, the theory for a unitary Fermi gas has
been extended to large mass-imbalanced systems~\cite{endo_quatriemes_2022}.
Large mass-imbalanced Fermi-Fermi mixtures can also be useful for quantum
simulation of the Fermi-surface effect~\cite{kondo_diffusion_1984,
kondo_diffusion_1984-1, kondo_muon_1986} which manifests itself in quantum
diffusion behaviors of heavy particles in solids~\cite{storchak_quantum_1998}.

We start with a brief introduction to the experiment and the experimental
procedures of relevance to the present research (Sec.~\ref{sec:experiment}).
Next, the results of our Feshbach resonance search for magnetic fields up to
$800\,\G$ are introduced (Sec.~\ref{sec:results}). We conclude with a
discussion of our findings and some thoughts on possible future
works~(Sec.~\ref{sec:discussion}).

\section{Experiment}
\label{sec:experiment}

\begin{figure*}[tb!]
	\centering
	\includegraphics{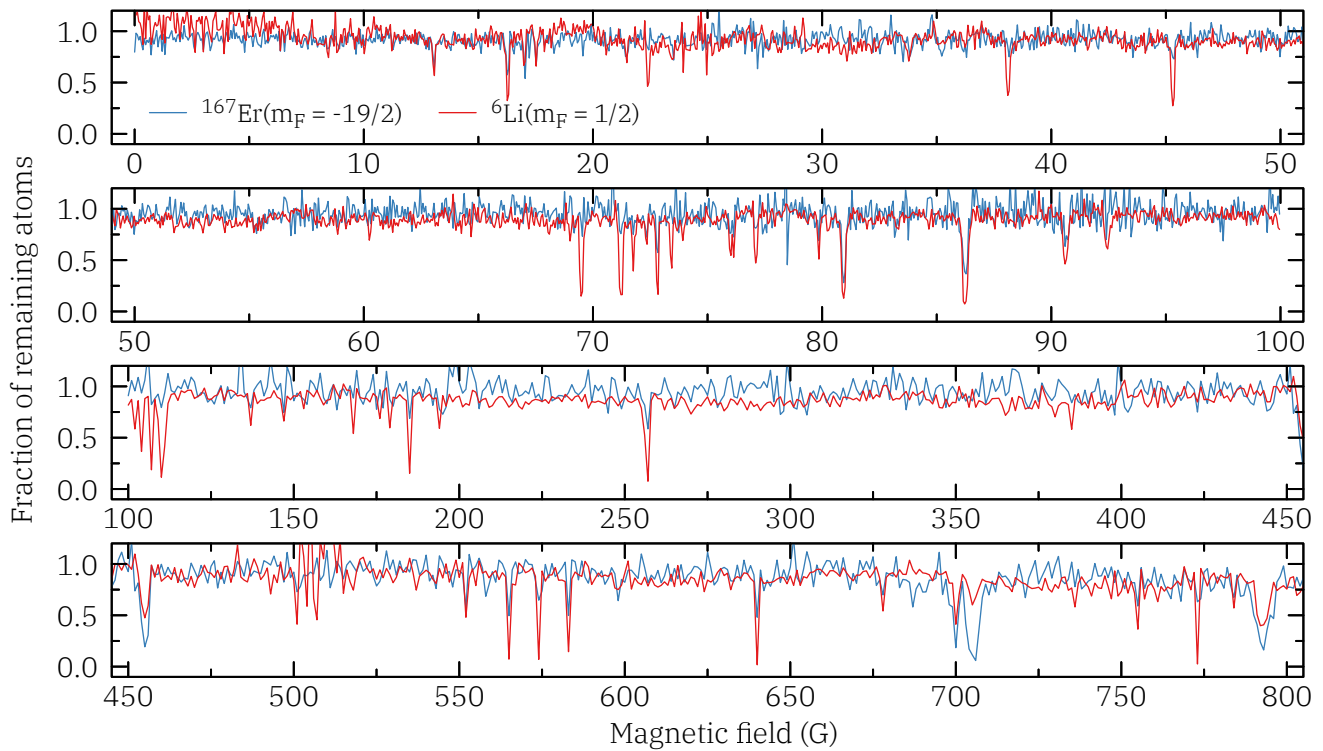}
	\caption{Magnetic field dependent atom loss in a \fEr($m_F = -19/2$)-\fLi($m_F
		= 1/2$) mixture. The panels show the fractions of Er atoms (blue) and Li
		atoms (red) that, after having been cooled to about $2$ to $3\,\uK$,
		remain in the optical trap after a holding time of $1000\,\ms$ at the
		various magnetic fields. The upper two panels cover the magnetic field
		range up to $100\,\G$ with data taken every $0.06\,\G$. The lower two
		panels encompass the range from $100$ to about $800\,\G$ at a resolution
		of $1\,\G$. (See the main text for a discussion on how the data was
		obtained and averaged.) Numerous resonant loss features are observed, most
		of them quite narrow at widths below $0.3\,\G$, but some broader
		resonances with widths above $1\,\G$ have also been identified.
	}
	\label{fig:fig1}
\end{figure*}

The experiment is based on the setup already used in our earlier
works~\cite{schafer_feshbach_2022}. However, while our previous efforts
focused on bosonic erbium, some upgrades to the machine have been necessary
for our present work with fermionic \fEr. This is because \fEr\ is known to
experience strong losses in optical traps operating at wavelengths around
$1064\,\nm$ for reasons that were never completely
uncovered~\cite{aikawa_reaching_2014}. An additional far-off resonant trap
(FORT) at $1550\,\nm$ is therefore prepared to circumvent this inconvenience.
Our modified experimental method is hence as follows: Starting as
in~\cite{schafer_feshbach_2022} a triple-species mixture at roughly $100\,\uK$
of \fEr, \fLi\ and \bYb\ is loaded into our horizontally oriented FORT
(H-FORT) operating at $1064\,\nm$. It is important to note here the addition
of bosonic ytterbium which we will use to sympathetically cool both Er and Li
in the evaporation step that is to follow. Since the magneto-optical trap
(MOT) light is blue-detuned for the Yb atoms in a FORT at $1550\,\nm$ and thus
causes considerable heating and atom loss we are required to first load all
atoms from the MOT into the $1064\,\nm$ H-FORT from where, once all MOT lights
could be extinguished, they are transferred within $30\,\ms$ into the
superimposed $1550\,\nm$ H-FORT. The transfer efficiency for each of the three
species is about 50\%. Forced evaporation proceeds in a crossed FORT
configuration where in addition to the H-FORT beam (waist diameter $50\,\um$)
a vertical FORT beam (V-FORT, waist diameter $240\,\um$) at the same
wavelength is added for a tighter confinement during the evaporation. As the
V-FORT beam is derived from the zeroth-order light of the acousto-optic
modulator that prepares the light for the H-FORT, the V-FORT is initially at
very low power. However, as the power of the H-FORT is gradually reduced
during the evaporation, the V-FORT power can be increased to efficiently
support the evaporation efficiency during most important final stages of the
evaporation. During evaporation the magnetic fields are carefully chosen for
good cooling performance and to maintain the spin polarization of the sample.
The magnetic field is initially set to $1.55\,\G$ and after $4\,\s$ reduced to
$0.4\,\G$. This procedure ensures that the natural spin-polarization of \fEr\
in the lowest $F = 19/2, m_F = -19/2$ magnetic sublevel obtained during the
narrow linewidth MOT is maintained. We further found that with this sequence
\fLi\ is also naturally polarized in its lowest magnetic $F = 1/2, m_F = 1/2$
state, alleviating the need for any active optical pumping. After evaporation
the remaining \bYb\ atoms are removed from the trap by a short pulse of light
resonant to the ${}^1S_0 \rightarrow {}^3P_1$ transition at $556\,\nm$ leaving
a pure Er-Li sample for the main part of the experiment.

In our first set of experiments we are interested in a general overview of the
\fEr-\fLi\ Feshbach resonance structure in the reasonable magnetic field range
of up to about $800\,\G$. For this we choose an evaporation ramp that has a
duration of $7\,\s$ and leaves the mixture in a trap with frequencies of about
$(\omega_x, \omega_y, \omega_z) = 2\pi \times (47, 405, 400)\,\Hz$ where the
$z$-axis is in vertical direction. After evaporation we typically obtain
$20(5) \times 10^3$ Er atoms at a temperature of $2.4(2)\,\uK$ and $7(2)
\times 10^3$ Li atoms at $3.0(5)\,\uK$. Next, the magnetic field is raised in
$10\,\ms$ to its desired value and the mixture is allowed to interact for
$1000\,\ms$ after which the magnetic field is again lowered and the number of
remaining atoms for both species is measured by standard absorption imaging.
For each magnetic field setting this experiment is repeated three times.
Additionally, at each field control measurements are taken in which either one
of the two species is removed from the optical trap by a pulse of resonant
$583\,\nm$ (Er) or $671\,\nm$ (Li) light as in~\cite{schafer_feshbach_2022}.
These control measurements are repeated two times each. From the
ratio of the averaged values we obtain our main observable: the fraction of
remaining atoms in the trap for each species. The results are summarized in
Fig.~\ref{fig:fig1}. The data can be divided into two parts: Up to $100\,\G$
the magnetic field was scanned with a resolution of $0.06\,\G$ to get a better
understanding of also the finer Feshbach resonance structure. The remainder of
the data has been taken with a step size of $1\,\G$. This greatly speeds up
the measurement process and allows us to obtain a good overview of the
available broader Feshbach resonances in a reasonable time. It is these
resonances that are most important for the physics of interest here.

\section{Results}
\label{sec:results}

\begin{table}[b!]
	\caption{List of identified \fEr($m_F = -19/2$)-\fLi($m_F = 1/2$)
		interspecies Feshbach resonances obtained from the data of
		Fig.~\ref{fig:fig1}. The resonance positions $B_0$ and widths $\Delta B$
		are from Lorentzian fits to the data. Note has to be taken of the two
		different resolution regimes: Up to $100\,\G$ more narrow resonances could
		be observed due to the higher measurement resolution whereas above that
		field purposely only broad resonances were detected and listed here. At
		$72.3\,\G$ the fit failed to provide a good resonance width estimated due
		to the finite measurement resolution.
	}
	\begin{tabular}{c@{\extracolsep{10mm}}crr@{\extracolsep{10mm}}c}
		\hline
		\hline
		& Resolution & $B_0$ ($\G$) & $\Delta B$ ($\G$) &\\
		\hline
		& $0.06\,\G$ & $13.1$ & $0.1$ &\\
		& & $16.3$ & $0.1$ &\\
		& & $17.0$ & $0.1$ &\\
		& & $17.5$ & $0.1$ &\\
		& & $22.4$ & $0.1$ &\\
		& & $33.8$ & $0.2$ &\\
		& & $38.1$ & $0.2$ &\\
		& & $45.3$ & $0.2$ &\\
		& & $69.5$ & $0.2$ &\\
		& & $71.2$ & $0.2$ &\\
		& & $71.8$ & $0.1$ &\\
		& & $72.3$ & --- &\\
		& & $72.8$ & $0.1$ &\\
		& & $73.4$ & $0.1$ &\\
		& & $76.1$ & $0.2$ &\\
		& & $77.1$ & $0.1$ &\\
		& & $79.9$ & $0.1$ &\\
		& & $80.9$ & $0.2$ &\\
		& & $86.2$ & $0.3$ &\\
		& & $90.6$ & $0.2$ &\\
		& & $92.5$ & $0.2$ &\\
		\hline
		& $1.0\,\G$ & $110.3$ & $1.5$ &\\
		& & $256.7$ & $1.0$ &\\
		& & $455.0$ & $2.3$ &\\
		& & $700.2$ & $1.2$ &\\
		& & $705.3$ & $2.6$ &\\
		& & $792.7$ & $3.5$ &\\
		\hline
		\hline
	\end{tabular}
	\label{tab:resonancelist}
\end{table}

Looking at the results in Fig.~\ref{fig:fig1} one immediately recognizes a
large number of resonant loss features. Focusing first on the upper two panels
that cover the magnetic field range up to $100\,\G$ we identify by eye $21$
resonances. From Lorentzian fits to these resonances we obtain a rough
estimate of their positions $B_0$ and full-widths at half maximum $\Delta B$.
All parameters are listed in the upper part of Tab.~\ref{tab:resonancelist}.
Note that due to our selected limited resolution of $0.06\,\G$ we do not
expect to have obtained a complete list of resonances. Also all the
uncertainties of $B_0$ and $\Delta B$ are estimated to be at least $0.1\,\G$.
Our primary goal with these first experiment has been to get a good overview
of typical resonance densities, on their widths and strengths. In this
respect, generally the spectrum is similarly rich as in our earlier results
with bosonic Er~\cite{schafer_feshbach_2022}. However, with all resonance
widths being well below $0.5\,\G$ it appears unlikely that the observed
resonances would be primary candidates to support the search of novel Efimov
and superfluid states.

We therefore now turn our attention to the lower two panels of
Fig.~\ref{fig:fig1} for the range $100$ to $800\,\G$. In this coarsely scanned
magnetic field range several narrow resonances are still visible where a
single data point happens to be close enough to such a resonance. We will
ignore such single-data-point events and instead focus on the broader
resonances of which in total $6$ have been observed. They are listed in the
lower part of Tab.~\ref{tab:resonancelist}. Of particular interest seem
the resonances at $455$, $705$ and $793\,\G$. Upon closer inspection one
notices that only the two resonances at $455$ and $793\,\G$ show nice losses
for both species. Of these two remaining resonances we find that the higher
one suffers from quite significant background losses also outside of the
immediate vicinity of the resonance. For these reasons we will now focus on
the resonant loss peak at about $455\,\G$ which additionally is also a
magnetic field strengths that is experimentally quite comfortable to work
with.

\begin{figure}[tb!]
	\centering
	\includegraphics{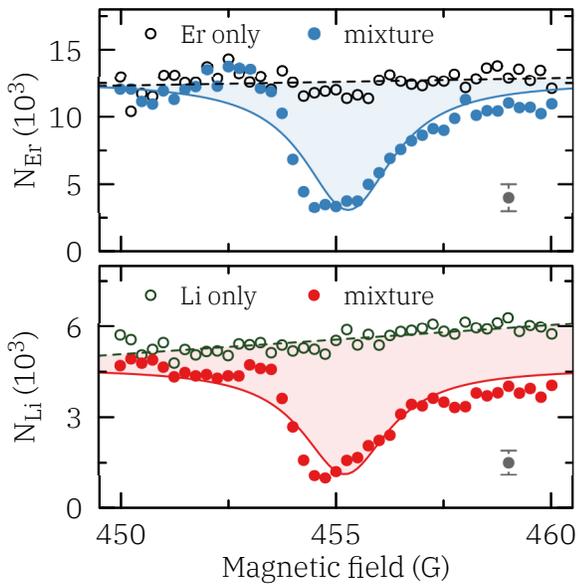}
	\caption{Results of a detailed measurement of the interspecies Feshbach
		resonance at about $455\,\G$. The temperature of the sample is about
		$1\,\uK$ lower than in the overview measurements of Fig.~\ref{fig:fig1}
		and the number of remaining atoms in the trap after only $30\,\ms$ of
		interaction time is shown. (See the main text for details.) A coordinated
		loss of both Er (top panel, blue points) and Li atoms (bottom panel, red
		points) in the mixture is observed while in the single species control
		measurements (open circles) no losses are found. The typical statistical
		error of the data is indicated in each panel by a single error bar example
		(gray). Lorentzian fits to the data on the one hand have their minima at
		$455.3\,\G$ and indicate widths of about $2.4\,\G$ but on the other hand
		also highlight the pronounced asymmetric lineshapes of the loss
		resonances.
	}
	\label{fig:fig2}
\end{figure}

For a more detailed view of this resonance we now work with a slightly
modified evaporation sequence: The ramp has been extended by $2\,\s$ to now
$9\,\s$ and the final trap frequencies are about $(\omega_x, \omega_y,
\omega_z) = 2\pi \times (49, 285, 272)\,\Hz$. After evaporation we then obtain
a colder sample with $13(2) \times 10^3$ Er atoms at a temperature of
$0.9(1)\,\uK$ and $5(1) \times 10^3$ Li atoms at $1.3(3)\,\uK$. The magnetic
field range from $450$ to $460\,\G$ is measured in steps of $0.25\,\G$. The
interaction time is reduced to $30\,\ms$. The data is otherwise taken and
averaged as before and we will this time directly look at the raw atom numbers
in Fig.~\ref{fig:fig2}. As expected, while in the mixture one observes a
nicely synchronized loss of both species close to resonance, there is no
magnetic field dependence in the single species data. The lineshapes of both
resonances are strongly asymmetric. This is particularly highlighted when
trying to describe them by Lorentzian fits (also included in
Fig.~\ref{fig:fig2}). Still, from the fit minima one can at least deduce the
approximate atom loss for both species. The reduction in the number of atoms
is $(9.5 \pm 1.4) \times 10^3$ for Er and $(4.5 \pm 0.6) \times 10^3$ for Li.
This implies that for every Li atom lost about $(2.1 \pm 0.4)$ Er atoms are
removed from the trap and is consistent with the assumption that the losses
are dominated by Er-Er-Li three-body collisions~\cite{ye_observation_2022}.

\begin{figure}[tb!]
	\centering
	\includegraphics{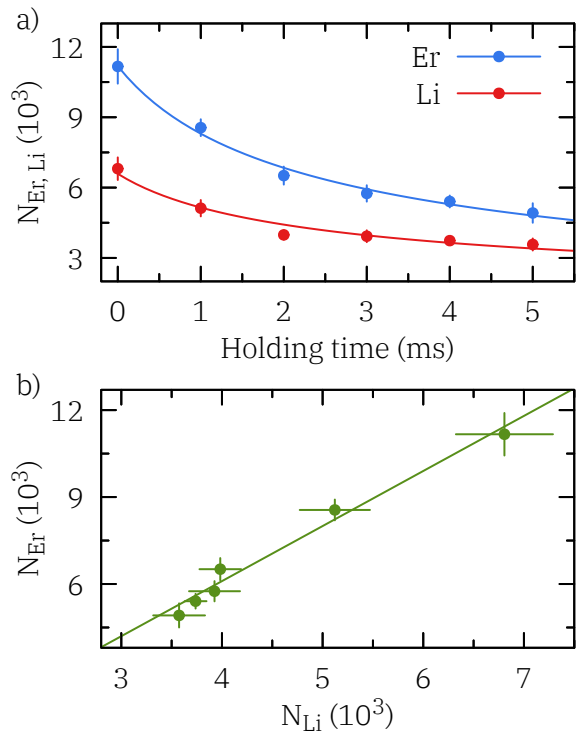}
	\caption{
		Decay dynamics of the Er-Li mixture at $455.0\,\G$. Panel a) shows
		the number of remaining Er and Li atoms (blue and red points) in the trap
		for holding times up to $5\,\ms$. The curves are fits to the data of an
		Er-Er-Li three-body loss model (see the main text for details). Panel b)
		indicates for each holding time the relationship between the remaining Er
		and Li atoms. The solid line is a fit of $N_{\rm Er} = a\,N_{\rm Li} + b$
		to the data with $a = 1.9(1)$ and $b = -1.5(6)\times10^3$.
	}
	\label{fig:fig3}
\end{figure}

In Fig.~\ref{fig:fig3} we take a closer look the typical decay dynamics by
studying the number of remaining atoms, $N_{\rm Er}$ and $N_{\rm Li}$, in the
trap at a magnetic field of $455.0\,\G$ and for holding times up to $5\,\ms$.
In Fig.~\ref{fig:fig3}a) the decay is described by a three-body decay
model~\cite{ye_observation_2022} (solid lines), again assuming Er-Er-Li
three-body collisions only, with a three-body collision rate coefficient $K_3
= 2(1) \times \lossunit$. Finally, Fig.~\ref{fig:fig3}b further corroborates
the applicability of such a model and also our earlier estimation of the
relative atom loss by directly showing the relationship between $N_{\rm Er}$
and $N_{\rm Li}$ which is well described by a linear fit of slope $1.9(1)$.
This is in good agreement with a complete loss of all particles from the trap
in Er-Er-Li three-body collisions which would lead to an expected slope of
$2.0$.

\section{Summary and Prospects}
\label{sec:discussion}

In the present work three major steps towards establishing ultracold
\fEr-\fLi\ mixtures as a new platform for the study of many-body physics in
general and Efimov states in particular have been taken: First, we could show
that by means of sympathetic cooling with \bYb\ this mixture can be brought to
microkelvin temperatures and that for the selected spin states no unexpected
losses occur at the employed magnetic fields. Second, with these minimum
requirements fulfilled we could demonstrate that the mixture indeed supports a
wealth of Feshbach resonances that, however, are still sufficiently separated
to not interfere with each other. Third, while most of the observed resonances
are quite narrow some of them feature widths of at least $2\,\G$. There, it
has further to be pointed out that currently we can only estimate the width of
the inelastic collisional loss resonance and it is known that the width of the
elastic part of the resonance might as well be
larger~\cite{ye_observation_2022}.

We have in particular focused on the broad resonance at about $455\,\G$.
There, coordinated losses in both channels have been found and the observed
atom number decay dynamics seems to support an Er-Er-Li loss channel for the
Feshbach resonance. It is exactly this condition of two heavy Fermions
interacting with a light particle that in particular motivated the current
work as it brings the study of a new type of Efimov three-body state within
reach. Finally, the shape of the observed resonance is highly asymmetric. This
might be purely caused by the complicated scattering physics
involved~\cite{fouche_quantitative_2019}. It might, however, also at least
partially be caused by additional losses from Efimov
states~\cite{pires_observation_2014, tung_geometric_2014}. More detailed
measurements will be required and are currently planned to clarify this
important question in this promising mixture system.

\section*{Acknowledgments}
This work was supported by the Grant-in-Aid for Scientific Research of JSPS
Grants No.\ JP17H06138, No.\ 18H05405, and No.\ 18H05228, JST CREST Grant No.\
JPMJCR1673 and the Impulsing Paradigm Change through Disruptive Technologies
(ImPACT) program by the Cabinet Office, Government of Japan, and MEXT Quantum
Leap Flagship Program (MEXT Q-LEAP) Grant No.\ JPMXS0118069021 and JST
Moonshot R\&D - MILLENNIA Program (Grant No.\ JPMJMS2269).

%\bibliographystyle{apsrev4-1}
%\bibliography{167Er6LiFeshbach}

\end{document}